\documentclass[10pt,a4paper,oneside]{article}
\usepackage[utf8]{inputenc}
\usepackage[T1]{fontenc}
\usepackage{amsmath}
\usepackage{amsfonts}
\usepackage{amssymb}
\usepackage[numbers]{natbib}
\usepackage[english]{babel}
\usepackage{graphicx}
\author{Diop Amadou}
\title{Analysevariabilite}
\oddsidemargin - 0.1 cm
\topmargin - 1.46cm

\begin{document}
	\title{\textbf{Analysis of the evolution of agro-climatic risks in a context of
			climate variability in the region of S\'{e}gou in Mali.}}
	\author{Diop Amadou$^{1}$ and Barro Diakarya$^{2}$\\$^{1}${\small Universit\'{e} de S\'{e}gou (R\'{e}publique du Mali).
			diopamadou01@yahoo.fr }\\$^{2}${\small Universit\'{e} Thomas Sankara (Burkina Faso). dbarro2@gmail.com}}
	\maketitle
	
	\begin{abstract}
		In the Sahel region the population depends largely on rain-fed agriculture. In West Africa in particular, climate models turn to be unable to capture some basic features of present-day climate variability. This study proposes a contribution to the analysis of the evolution of agro-climatic risks in a context of climate variability. Some statistical tests are used on the main variables of the rainy season to determine the trends and the variabilities described by the data series. Thus, the paper provides a statistical modeling of the different agro-climatic risks while the seasonal variability of agro-climatic parameters were analized as well as their inter annual variability. The study identifies the probability distributions of agroclimatic risks and the characterization of the rainy season was clarified.
		
		
		
	\end{abstract}
	\section{Introduction}
	
	Climate change is at the center of current concerns. Indeed, these consequences directly impact the socio-economic development of all countries. In particular, in West Africa where ecosystems are fragile \citep{solomon2007climate}, climatic changes, combined with anthropogenic pressure, lead to significant ecological imbalances \citep{benoit2008changements,assemian2013etude}. For the IPCC, 2007, the continent's vulnerability to climate change would be much greater. Thus, an overall negative effect on the production of millet, maize and sorghum could occur \citep{sultan2012global,berg2013projections}, with critical thresholds of food insecurity in many regions	\citep{Roudier2012}. Frequent droughts, longer dry spells, land degradation and significant loss of biodiversity are also predicted among others.
	
	These are the challenges facing West African agriculture that is mainly rainfed and therefore very sensitive to climatic variability. In addition, as the main lever of national economies, its peasant practice is still based on production systems with performances subject to climatic hazards \citep{bazzaz1997changements}. In addition, this vulnerability was accentuated at the end of the 20th century by a decrease in rainfall and an increase in the population\citep{sultan2012global}. Indeed, a change of regime in the rainfall series is observed over the period 1968-1990 \citep{Nicholson2001,leBarbe2002rainfall,Lamb2006}. The droughts of the 1970s and 1980s also caused a sharp decrease in rainfall of 20 to 40\% between 1931-1960 and 1968-1990 in the Sahel against 15\% in the tropical and humid regions. A southward movement of isohyets of the order of 150 to 200 km is also observed in the Sudano-Sahelian zone \citep{diouf2001lutte}. According to \citep{Sivakumar1992,sultan2005agricultural}, the decrease in rainfall amounts can explain a 35 to 45\% drop in crop yields in the Sahel.
	
	These trends greatly impact the variability of the characteristics of the agricultural season. Indeed, one of the characteristics of the Sudano-Sahelian climate is the existence of a close and linear relationship between the date of onset of rains and the length of the cropping period \citep{sivakumar1988predicting}. The variation in the onset and cessetion dates of the rainy season, causing a fluctuation in the length of the rainy season resulting in delays and premature stops of the rains.
	
	One of the risks associated with these modifications is the difficulty of adapting plants to the local ecological disturbances induced. Particularly for certain species with relatively long reproduction cycles \citep{moorcroft2006potential}. Thus, a good characterization of the onset and cessetion dates of the season is essential to optimize the results of a cropping season.
	
	The main objective of this study is to analyze the temporal evolution of agro-climatic risks in a context of climatic variability in the region of S\'{e}gou in Mali. To this end, the paper is organized as follows. The
	section 2 describes the study area. The section 3 is concerned with an
	overview of the main statistical tests used in the study. Our results sing
	at section 4. We use tests on variables characterizing the rainy season to
	identify the trends and variabilities contained in the data series. Then, we
	propose a statistical modeling of these different agro-climatic risks. The
	data analysis techniques and tools retained for this study are presented
	below. \ The results on breaks in the series, the evolution of seasonal and
	intra-seasonal agro-climatic risks and their possible consequences are then
	discussed. 
	
	\section{Study area}
	\bigskip
	 Located in central Mali, the S\'{e}gou region 13${{}^\circ}$22$^{\prime}$05$^{\prime\prime}$nord, 5${{}^\circ}$16$^{\prime}$2$^{\prime\prime}$ouest covers an area of 64,947 km$^{2}$ (approximately 5\% of Mali). It is bounded to the south by the Sikasso region, to the south-east by Burkina Faso, to the east by the Mopti region to the north by Mauritania and the Timbuktu region and to the west by the Koulikoro region.\
	With the regions of Mopti and Koulikoro, it forms what is commonly referred to as the Center of Mali. It is mainly located in the Sahelian zone where it enjoys a semi-arid climate (average annual rainfall: 513 mm). The presence of several rivers (it is crossed by the Niger river (over 292 km) as well as the Bani river) allows irrigated crops. Indeed, the area is home to the \textit{Office du Niger} and the \textit{Office riz de S\'{e}gou,} which have large developed and irrigated areas.
	The S\'{e}gou region has 16 classified forests covering an area of 78.860 ha.
\begin{center}
	\includegraphics[width=3.9358in, height=3.7144in]{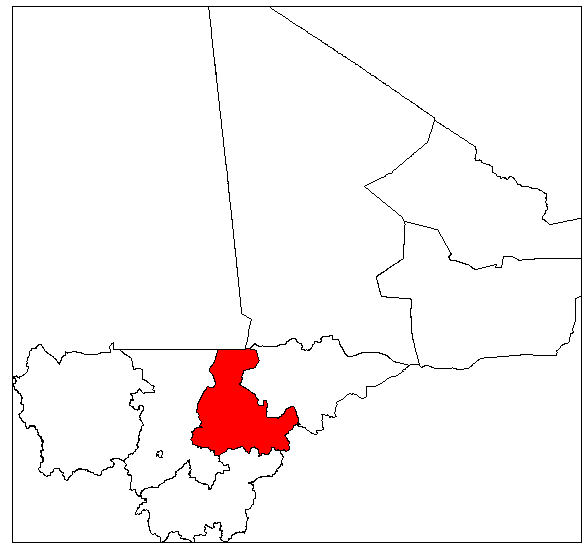} 
	
	Figure 1: S\'{e}gou region location in Mali (source : the author)
	\label{Segou region location }
\end{center}
\section{Methods}

In this study, daily data for around thirty years (1990-2019) from the
synoptic station of S\'{e}gou Latitude: $13{{}^\circ} 24^{\prime}N;$ Longitude: $06{{}^\circ} 09^{\prime}N$; Altitude of the station: $288.05 m$ on the variables of temperature, humidity, rains, insolation and wind speed were used. The missing
data were estimated and replaced by the daily averages obtained over the same period.

Initially, a set of graphical representations was used to visualize the trends of the various climatic parameters and the temperature and rainfall anomalies. The Mann-Kendall test was implemented to measure the significance of the trends observed on the different variables and the Pettitt test to identify the different break dates in the data. A break is characterized by a change in the probability law of the time series at a given instant \citep{lubes1994caracterisation}. Agronomic criteria were used to determine the onset and cessetion dates of
the season (Sivakumar and PRESAO). Correlation analyzes were also carried out between the different agro-climatic parameters.

\begin{itemize}
	\item \textbf{Mann-Kendall test}
\end{itemize}

The Mann-Kendall test \citep{mann1945nonparametric} has been widely used to detect trends in meteorological, hydrological and
agro-meteorological time series. It is used to determine with a nonparametric test whether a trend is identifiable in a time series that possibly includes a seasonal component.
The null hypothesis $H_{0}$ of these tests is that there is no trend. The three alternative hypotheses of negative, non-zero or positive trend can be chosen.

Mann-Kendall tests are based on the calculation of Kendall's tau measuring the association between two samples and itself based on the ranks within the samples.

Consider a dataset consisting of $x$ values with sample size $N$. The Mann-Kendall test (MK) calculation begins by estimating the $S$ statistic:

\begin{equation}
S={\displaystyle\sum\limits_{i=1}^{N-1}}{\displaystyle\sum\limits_{j=i+1}^{N}}sgn\left(  x_{j}-x_{i}\right)  \text{ \ \ \ }for\text{ }j>I
\end{equation}
KENDALL and STUART (1967); MANN (1945) states that when $N\geq8$, the distribution of $S$ approaches the Gaussian form with mean $E(S)=0$ and
variance $V(S)$ given by:
\begin{equation}
V\left(  S\right)  =\frac{N\left(  N-1\right)  \left(  2N+5\right) 	{\displaystyle\sum\limits_{m=1}^{SS}}t_{i}\left(  m-1\right)  \left(  2m+5\right)  m}{18}%
\end{equation}
Where $SS$ is the number of tied groups and $t_{i}$ is the length of the $SS^{th}$ group.

The statistic $S$ is then standardized (MK), and its significance is estimated from the normal cumulative distribution function.

\begin{equation}
MK=\left\{
\begin{array}
[c]{c}
\frac{S-1}{\sqrt{V\left(  S\right)  }}\rightarrow S>0\\
0\rightarrow S=0\\
\frac{S+1}{\sqrt{V\left(  S\right)  }}\rightarrow S<0
\end{array}
\right.
\end{equation}

\begin{itemize}
	\item \textbf{Pettitt test}
\end{itemize}

The Pettitt test is non-parametric and is derived from the Mann-Withney test one. The absence of a break in the series $\left(  x_{i}\right)  $ of size $N$constitutes the null hypothesis.

Pettitt defines the variable $U_{t,N}$:
\begin{equation}
U_{t,N}={\displaystyle\sum\limits_{i=1}^{t}}{\displaystyle\sum\limits_{j=t+1}^{N}}D_{i,j}
\end{equation}

Where: $D_{i,j}=sgn\left(  x_{i}-x_{j}\right)  $ with $sgn\left(  Z\right)=1$ if $Z>0,$ $0$ if $Z=0$ and $-1$ if $Z<0.$

He proposes to test the null hypothesis using the $K_{N}$ statistic defined by the maximum in absolute value of $U_{t,N}$ for $t$ varying from $1$ to $N-1$
From rank theory, Pettitt shows that if $k$ denotes the value of $K_{N}$ taken by the series studied, under the null hypothesis, the probability of exceeding $k$ is given approximately by:
\begin{equation}
\Pr ob\left(  K_{N}>k\right)  \approx2\exp\left[  \frac{-6k^{2}}{\left(N^{3}+N^{2}\right)  }\right]
\end{equation}
For a given first-kind $\alpha$ risk, if the estimated probability is less than $\alpha$, the null hypothesis is rejected. The series then includes a localized break at the moment $\tau$ when $K_{N}$ is observed.

\begin{itemize}
	\item \textbf{Sivakumar criterion}
\end{itemize}
The agronomic criterion proposed by \citep{sivakumar1988predicting} was retained to determine the dates of the start of the season. Indeed,
Sivakumar considers the onset of the rainy season in the Sahelian and Sudanese regions as the date from May 01 collecting a water depth of at least 20 mm over 3 consecutive days, without any dry sequences of more than 7 days in the following 30 days. The date of the end of the season is the day when, after September 01, there is no more rain greater than or equal to 5 mm for at least 20 successive days or two weeks.

\begin{itemize}
	\item \textbf{PRESAO criterion for cessetion date of the season}
\end{itemize}

A comparison of the cessetion dates of season was carried out using the criterion retained in the PRESAO\footnote{PREvision Saisonni\`{e}re en Afrique de l'Ouest} (Seasonal PREvision in West Africa) process set up in 1998 by the ACMAD\footnote{African Centre of Meteorological Application for Development}-AGRHYMET\footnote{AGRom\'{e}t\'{e}orologie-HYdrologie-MET\'{e}orologie}-ABN\footnote{Autorit\'{e} du Bassin du Niger}-ICRISAT\footnote{International Crops Research Institute for the \par Semi-Arid Tropics} consortium. The end of the season date for this criterion is when from 01 September, soil capable of holding 70 mm of available water is
completely taken up by a daily evapotranspiration loss of 5 mm.
\bigskip
For the fit of data to a normal distribution, we applied a Shapiro-Wilk test, associated with a quantile-quantile plot.

\begin{itemize}
	\item \textbf{Shapiro-Wilk test}
\end{itemize}
Published in 1965 by Samuel Sanford Shapiro and Martin Wilk, the Shapiro-Wilk test tests the null hypothesis that a sample $x_{1},x_{2},...x_{n}$ is from a normally distributed population.

The $W$ test statistic is defined by:%

\begin{equation}
W=\frac{\left(	{\displaystyle\sum\limits_{i=1}^{n}} a_{i}x_{\left(  i\right)  }\right)  ^{2}}{{\displaystyle\sum\limits_{i=1}^{n}}\left(  x_{i}-\overline{x}\right)  ^{2}}
\end{equation}
Where

\begin{itemize}
	\item $x_{\left(  i\right)  }$ denote the $i^{th}$ smallest number in the sample
	
	\item $\overline{x}$ is the sample mean
\end{itemize}

$\left(  a_{1},a_{2},...a_{n}\right)  =\frac{m^{T}V^{-1}}{\left(  m^{T}V^{-1}V^{-1}m\right)  ^{1/2}}$

Where $m=\left(  m_{1},m_{2},...m_{n}\right)  ^{T}$ are the expectations of the order statistics of iid variables according to a normal distribution and $V$ is the variance-covariance matrix of these order statistics.

If the p-value is greater than the chosen alpha level, then we accept the null hypothesis.

\begin{itemize}
	\item \textbf{The quantile-quantile plot (Q-Q plot)} is a graphical tool that makes it possible to assess the relevance of the fit of a given distribution to a theoretical model. From the observed statistical series, we then calculate a certain number of quantiles $x_{i}$. If the statistical series follows the chosen theoretical distribution well, we should have the observed quantiles $x_{i}$ equal to the quantiles $x_{i}^{\ast}$ associated with the theoretical model.
\end{itemize}
	
	\section{Results}
	
	\bigskip
	
	\subsection{Seasonal variability of agro-climatic parameters}
	
	\subsubsection{Seasonal temperature variability}
	
	\bigskip Analysis of the seasonal evolution of minimum, maximum and average temperatures (Figure 2) over the period 1990-2019 shows a ripple throughout the year. Minimum temperatures present maximums in April-May with an average of 27.56${{}^\circ}$C for the month of May and minimums in the months of January and December with an average of 17.26${{}^\circ}$C for the month of January. These low temperatures would certainly be due to the cool and dry Harmattan winds coming from the Sahara and sweeping the entire West African sub-region. The months of July, August and September show a very low dispersion of minimum temperatures which correspond to the winter period of the area. Maximum temperatures also show two peaks; the first observed during the months of March-April-May with temperatures reaching an
	average of 40${{}^\circ}$C and the second during the months of October and November. The months of July, August and September are the cooler months with temperatures in the range of 30-36${{}^\circ}$C. The period of March-April-May corresponds to the dry season, which
	explains the high temperatures observed. The average temperatures vary between 25 and 28${{}^\circ}$C for the minimums and 30 and 34${{}^\circ}$C for the maximums.
	
	\bigskip

	\begin{center}
		\includegraphics[height=3.3537in,width=4.6942in]
		{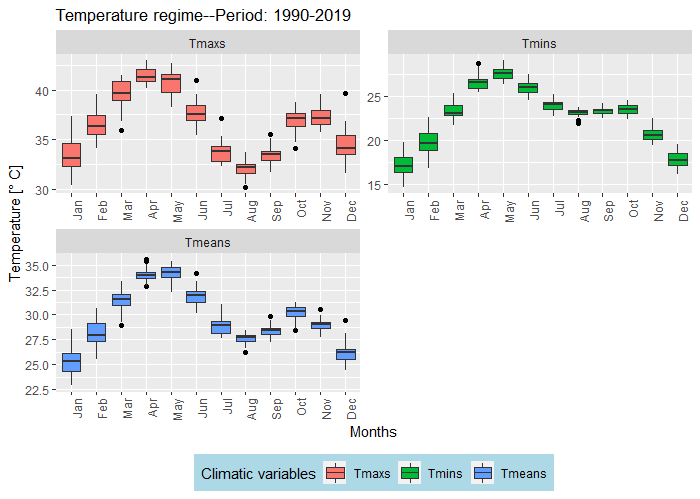}
		
		Figure 2: Temperature regime
		\label{Temperature regime}
	\end{center}
	
	\subsubsection{Seasonal rainfall and evapotranspiration variability}
	\bigskip 
	Rainfall regime in the area is uni modal (Figure 3). The monthly average rainfall shows a high variability over the year. The period from June to September is the wettest with quantities on average over 100 mm. August is the wettest month with average heights reaching 200 mm. However, the monthly total shows a strong dispersion during this month. Evapotranspiration regime seems to draw the opposite of the trend of that of rain
	From this analysis, we can say that the rainy season in the area is from June to September with more than 80\% of the year to date.
	
	\bigskip

	\begin{center}
		\includegraphics[
		width=5.015in
		]
		{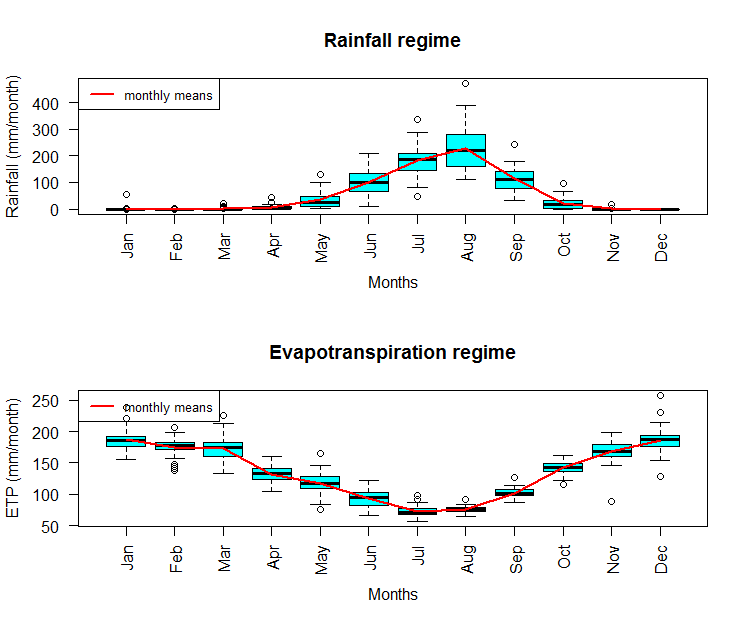}
	
		Figure 3: Rainfall and ETP regime
		\label{Rainfall and ETP regime}
	\end{center}
	Daily precipitation levels show (Figure 4) very variable frequencies over the study period (1990-2019). Daily heights of less than 5 mm are very frequently observed in the area. However, precipitation greater than 40 mm although very rare was also observed.

	\begin{center}
		\includegraphics[
		height=3.3537in,
		width=4.6942in
		]
		{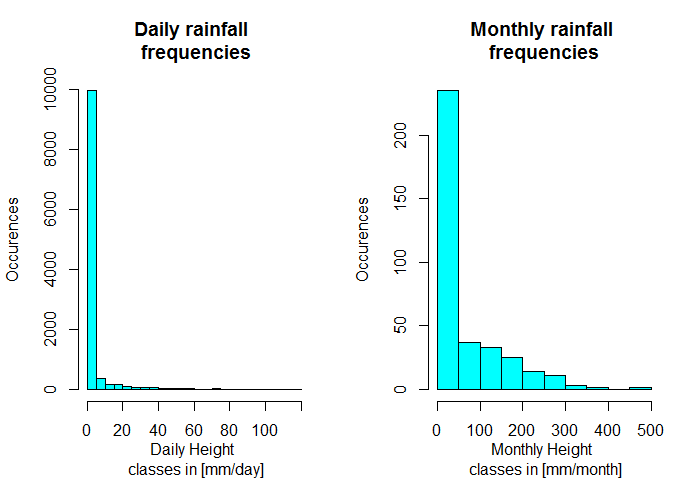}%
		
		Figure 4 : Rainfall occurancy
		\label{Rainfall occurancy}%
	\end{center}

	\subsubsection{Seasonal humidity variability}
	
	\bigskip
	 Change in relative humidity over the season (Figure 5) is shown as an asymmetric bell curve to the right. Minimum humidity is below 25\% during the months of November to April and is above 40\% between July and September. Maximum humidity has the same evolution as the minimum with rates below 60\% in the dry season and above 80\% between July and September. However, it should be noted that the two types of humidity behave in an opposite way in terms of dispersion. In fact, the minimum humidity has a strong dispersion in the wet season and a weak dispersion between October and April, unlike the maximum humidity which shows a weak dispersion between July and October and a strong dispersion the rest of the year.
	
	\begin{center}
		\includegraphics[
		height=3.3537in,
		width=4.6942in
		]
		{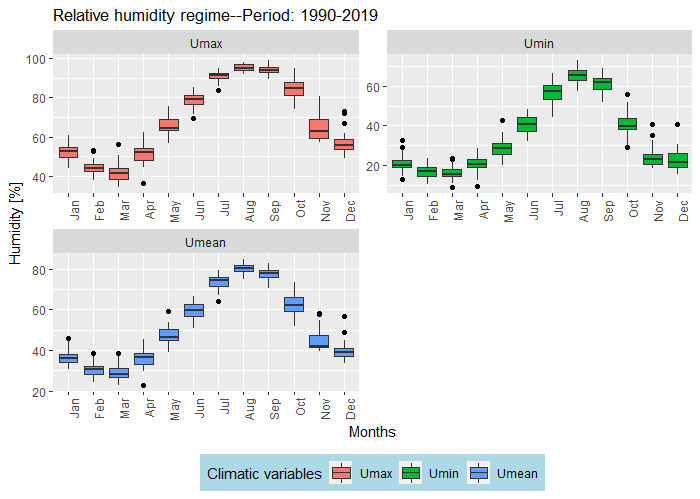}
	
		Figure 5 : Relative humidity regime
		\label{Relative humidity regime}
	\end{center}

	\subsubsection{Seasonal wind speed variability}
	
	\bigskip
	 Wind speed is generally between 0.5 m/s and 2.5 m/s (Figure 6). Strong winds are recorded during the months of January, February, May and June with a speed between 1.5 m/s and 2.5 m/s. This period also records a strong dispersion of the wind speed which could be explained by the installation of the rainy season with the predominance of thunderstorms for the months of May and June and the cold season which extends from December to February.
	\begin{center}
		\includegraphics[
		height=3.0537in,
		width=4.6942in
		]
		{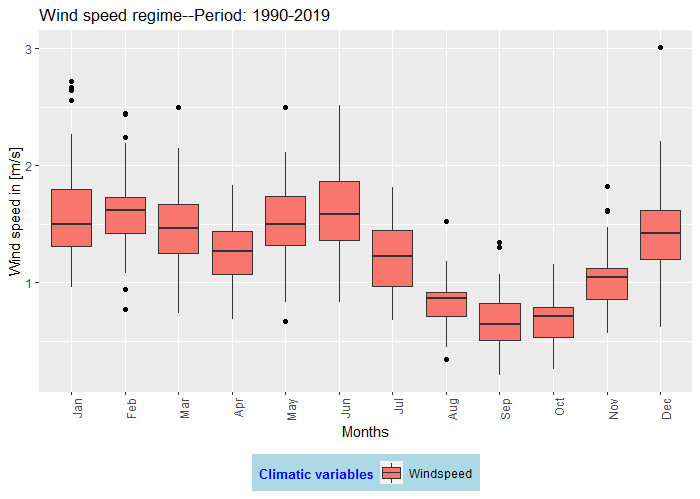}
	
		Figure 6 : Wind speed regime
		\label{Wind speed regime}
	\end{center}
	The strong monthly wind speed occurrences are winds below 2 m/s. Strong winds of 2 to 3 m/s are less frequent in the area (Figure 7).
	\begin{center}
		\includegraphics[
		height=3.3537in,
		width=4.6942in
		]
		{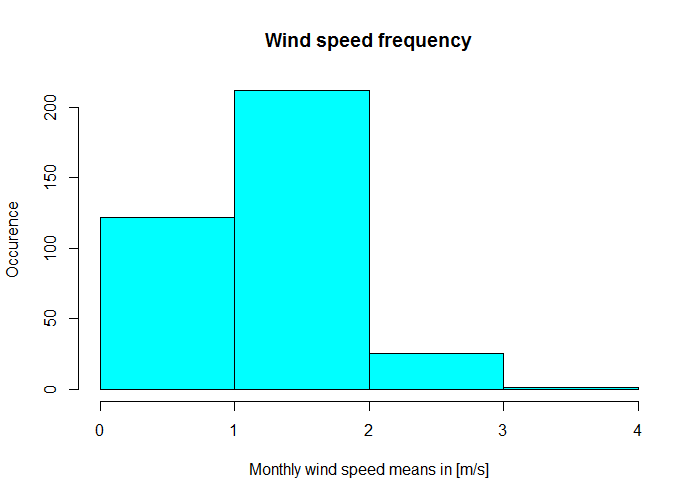}

		Figure 7 : Wind speed occurancy
		\label{Wind speed occurancy}
	\end{center}
	\subsubsection{Seasonal insolation variability}
	
	\bigskip 
	Monthly average insolation and its monthly cumulative (Figure 8) experience the same type of dispersion during the year and over the period 1990-2019. In fact, there is little dispersion during the wet season and great dispersion the rest of the year. Monthly averages vary between 6 and 9 hours and the cumulatives are generally greater than 225 hours per month. The smallest averages are observed during the months of July, August and September due to the frequency of cloud cover during the rainy season which prevents
	solar radiation from reaching the earth's surface and the strongest in the dry season.
	\begin{center}
		\includegraphics[
		height=3.3537in,
		width=4.6942in
		]
		{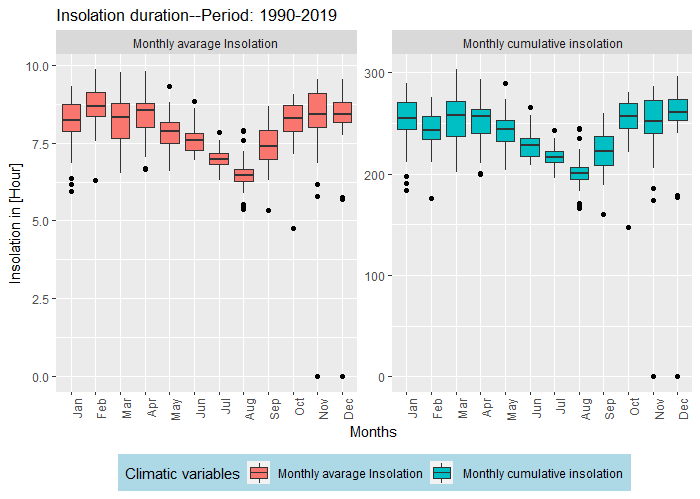}

		Figure 8 : Insolation regime
		\label{Insolation regime}
	\end{center}

	\subsection{Inter annual variability of agro-climatic parameters}
	
	\subsubsection{Inter annual temperature variability}
	
	\bigskip 
	Inter annual variation of temperatures (Figure 9) shows an increasing trend for both minimum and maximum. However, the increase is greater but not significant (Mann-Kendal test used with a significance level at the 5\% threshold) for minimum temperatures (0.14${{}^\circ}$C / 10 years) than maximum temperatures (0.12${{}^\circ}$C /10 years). The Pettitt test at the 5\% threshold made it possible to detect the years of ruptures. For the minimum temperatures, the year 2001 was detected as the year of rupture with a p-value$<$0.05 on the other hand there is no significant rupture for the maximum temperatures. The difference between the rupture periods (before 2001 and after 2001) is of the order of 0.4${{}^\circ}$C. Temperature rise becomes continuous and more significant on minimum temperatures than on maximum temperatures.
	\begin{center}
		\includegraphics[
		height=3.8891in,
		width=5.175in
		]
		{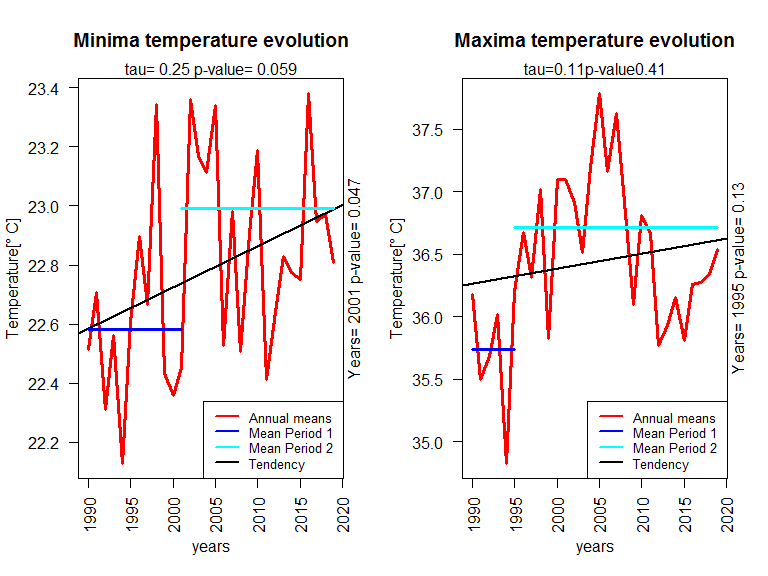}

		Figure 9: Temperature inter annual variabilities
		\label{Temperature variabilities}
	\end{center}

	\begin{center}
		\includegraphics[
		height=3.8891in,
		width=5.175in
		]
		{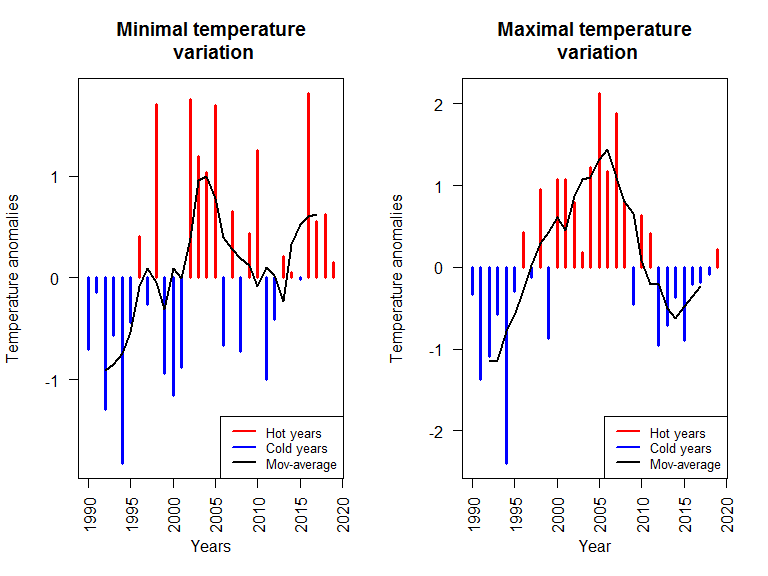}
		
		Figure 10 : Temperature anomalies
		\label{Temperature anomalies }
	\end{center}
	Overall, the analysis of minimum, maximum and average temperature anomalies (Figures 9;10) shows a succession of hot decades (1990-1999 and 2010-2019) and cold decades (2000-2009) throughout the study period. If this trend continues, we could see more warm years in the next 10 years.
	
	\begin{center}
		\includegraphics[
		height=3.8891in,
		width=5.175in
		]
		{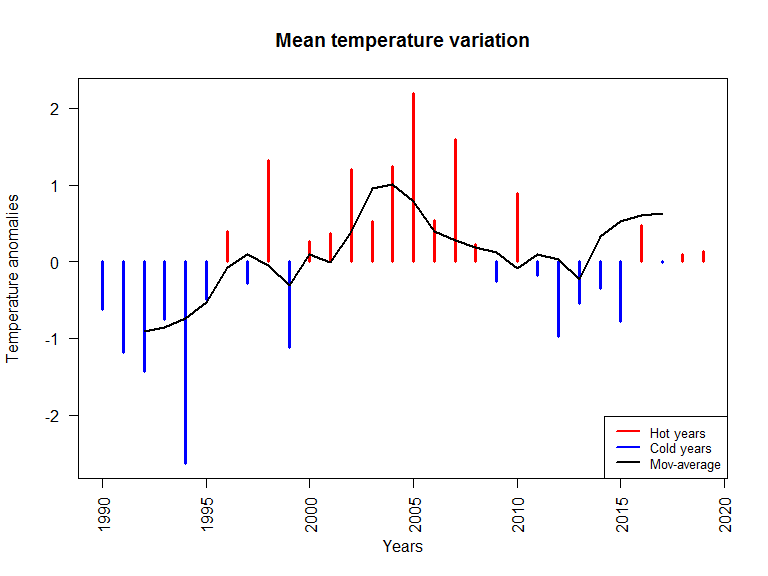}

		Figure 11: Mean temperature anomalies
		\label{Mean temperature anomalies}
	\end{center}
	\subsubsection{Inter annual rainfall variability}
	
	\bigskip
	 The evolution of annual cumulative and annual maximum rainfall (Figure 12) is marked by a high variability in cumulatives (500- and over 1000 mm) with an increasing trend which, however, is not significant in the Mann-Kendall test. Also the Pettitt test was found to be insignificant with p-values {}{}$>$0.05. Nevertheless the difference in average calculated between rupture periods is worth approximately 75 mm. In other words, the last 10 years have seen a drop in rainfall amounts of around 75 mm. In the distribution of annual maximum rainfall, 50\% of the rainfall is greater than 65 mm and it exceeds 73 mm every 8 out of 30 years.
	\begin{center}
		\includegraphics[
		height=3.8891in,
		width=5.175in
		]
		{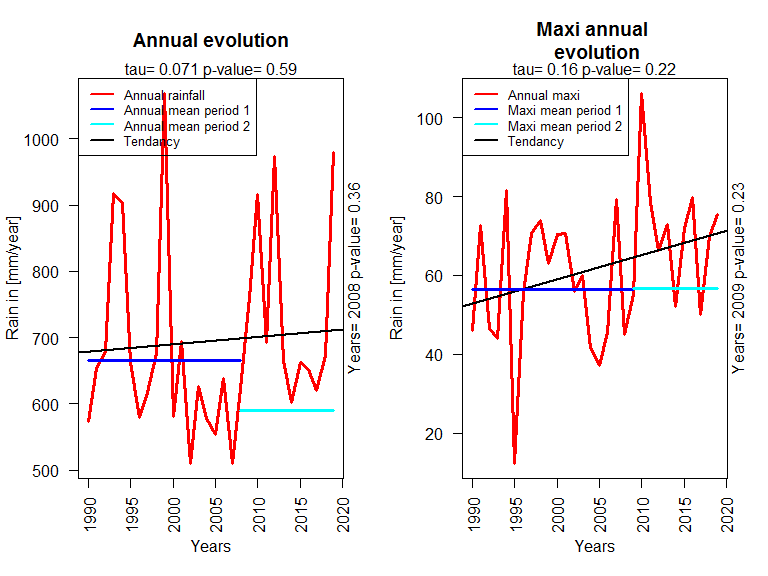}
		
		Figure 12: Rainfall variability
		\label{Rainfall variability}
	\end{center}
	\bigskip 
	The analysis of the variability of cumulative rainfall (Figure 13) shows a marked trend of years in deficit over the period 2001-2008 and two periods of excess years: a shorter one from 1992 to 1995 and a longer one from 2009 to 2019. The remarkable deficit years are 2002 and 2007.
	\bigskip
	\begin{center}
		\includegraphics[
		height=3.8891in,
		width=5.175in
		]
		{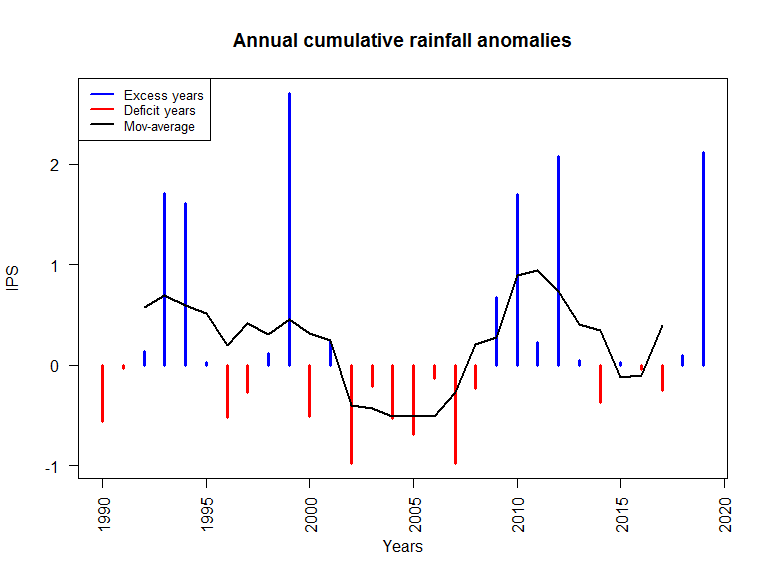}
		
		Figure 13: Cumulative rainfall variability
		\label{Cumulative rainfall variability}
	\end{center}
	\subsubsection{Inter annual wind speed variability}
	
	\bigskip 
	Annual wind speed averages are between 0.5 and 2 m / s (Figure 14). They show a downward trend. Maximum daily wind speed exceeds in more than 75\% of the observations the 04 m / s.
	\begin{center}
		\includegraphics[
		height=3.8891in,
		width=5.175in
		]
		{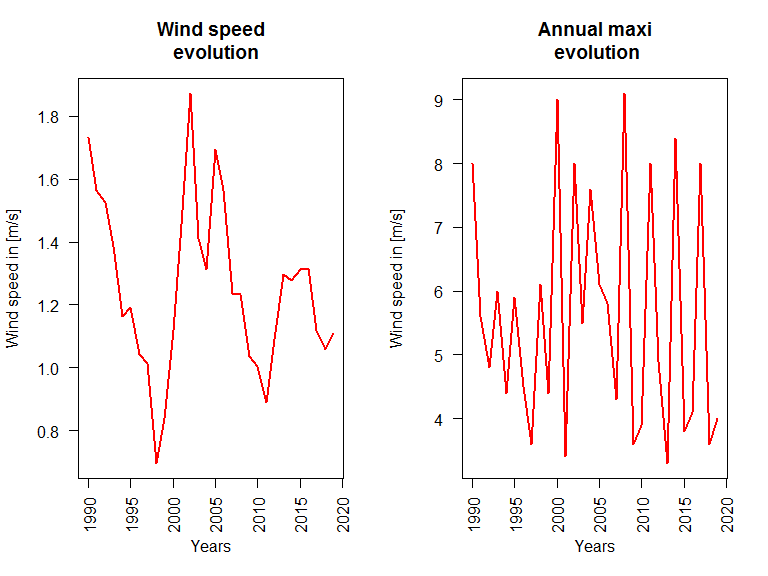}
		
		Figure 14: Inter annual wind speed evolution
		\label{Inter annual wind speed evolution}%
	\end{center}
	
	\bigskip
	
	\subsubsection{Inter annual insolation and humidity variability}
	
	\bigskip 
	Insolation shows a decreasing trend and varies between 6 and 8.5 hours on average per year (Figure 15). Relative humidity shows low values {}{}for 2002 and 2007, which correspond to deficit years .
	\begin{center}
		\includegraphics[
		height=3.8891in,
		width=5.175in
		]
		{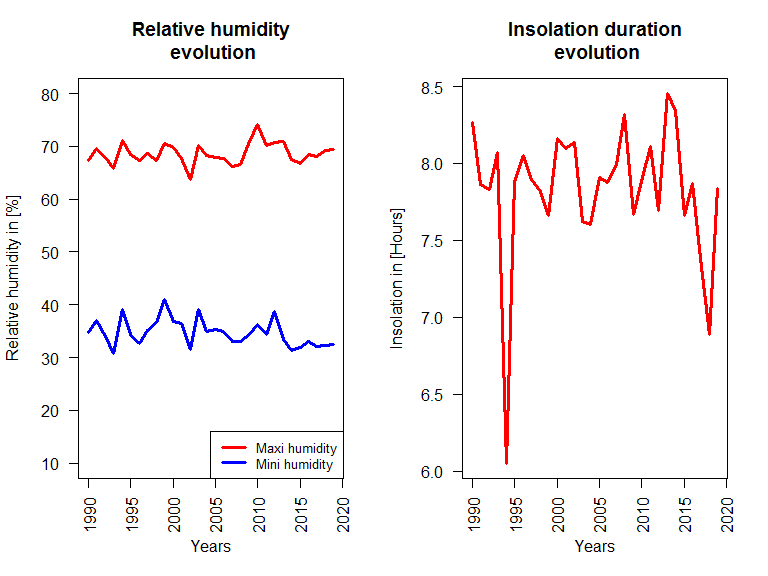}
		
		Figure 15: Insolation and humidity evolution
		\label{Insolation and humidity evolution}
	\end{center}
	\subsubsection{Inter annual evapotranspiration variability}
	
	\bigskip
	 The overall trend of evapotranspiration calculated from the Penman-Monteith formula is slightly downward over the period and despite a high variability observed, the trend and break tests were not significant
	(Figure 16). However, the difference observed between the two averages of the two periods is of the order of 0.22 mm.
	\begin{center}
		\includegraphics[
		height=3.8891in,
		width=5.175in
		]
		{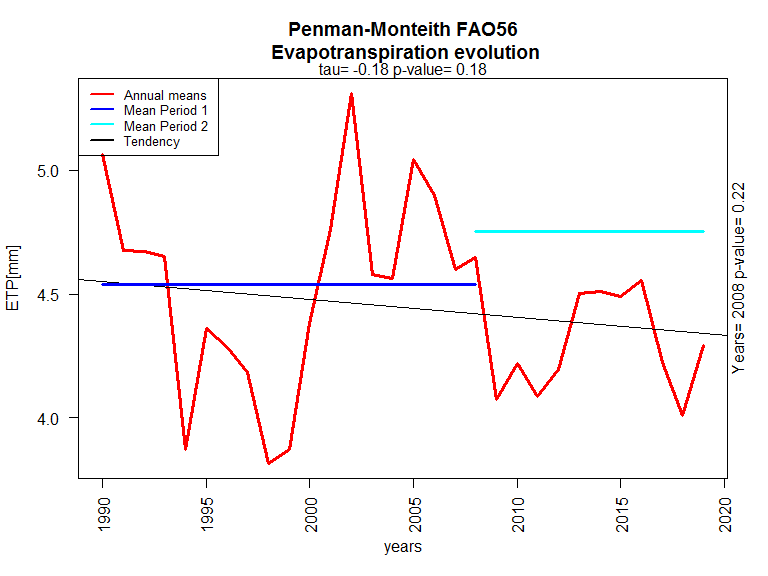}

		Figure 16: ETP inter annual evolution
		\label{ETP inter annual evolution}
	\end{center}
	
	\subsection{Characterization of the rainy season}
	
	\subsubsection{Analysis of the onset date of the rainy season}
	
	\bigskip 
	The distribution of the onset dates of the season is symmetrical and presents a large dispersion and a very high variability (Figure 17). They take place between April 12 at the earliest and June 18 at the latest. The average onset date of the season is around 07 June. In 75\% of cases, the onset dates
	are less than June 15. The farmer runs less risk of re-sowing by sowing after this date because it is observed 3 years out of 4. Also any farmer who sows before May 15th runs a great risk of re-sowing because its occurrence is 1 year out of 4 .
	\begin{center}
		\includegraphics[
		height=3.8891in,
		width=5.175in
		]
		{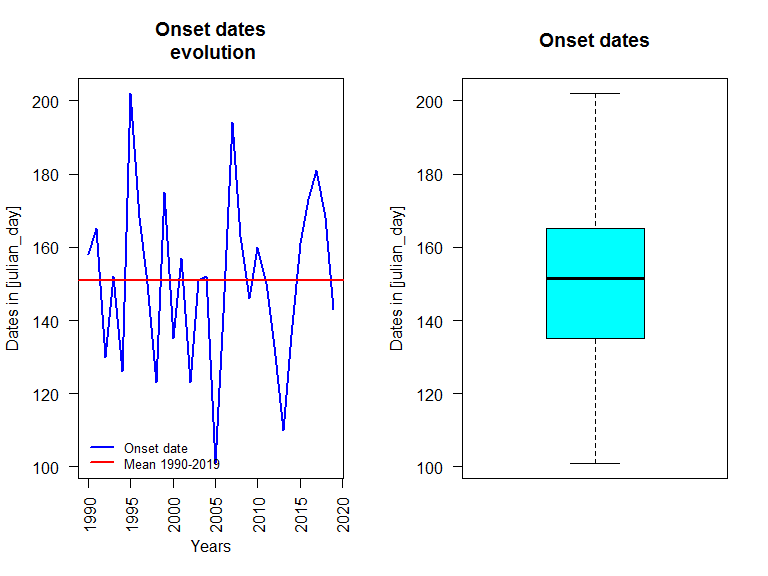}
	
		Figure 17: Onset dates evolution
		\label{Onset dates evolution}
	\end{center}
	The late start of the season dates are marked by high variability over the entire study period and remain strong from 2015 onwards (Figure 18).
	\begin{center}
		\includegraphics[
		height=3.8891in,
		width=5.175in
		]
		{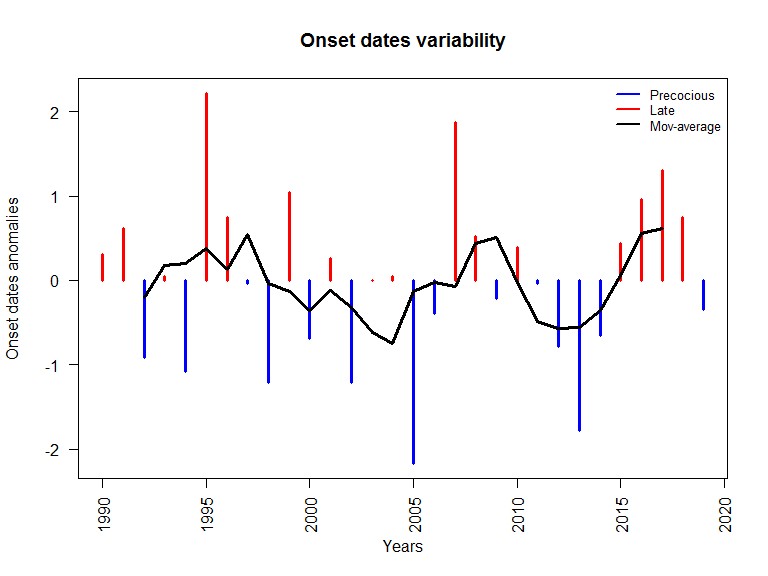}
	
		Figure 18: Onset dates anomalies
		\label{Onset dates anomalies}
	\end{center}
	
	\subsubsection{Analysis of the cessetion dates of the rainy season}
	\bigskip 
	The cessetion dates of season vary between September 10 and October 20 for the Sivakumar criterion and between September 2 and October 17 for the PRESAO criterion(Figure 19). In addition, the two criteria meet on average with dates around October 30 for Sivakumar and October 28 for PRESAO.
	\begin{center}
		\includegraphics[
		height=4.0811in,
		width=5.2474in
		]
		{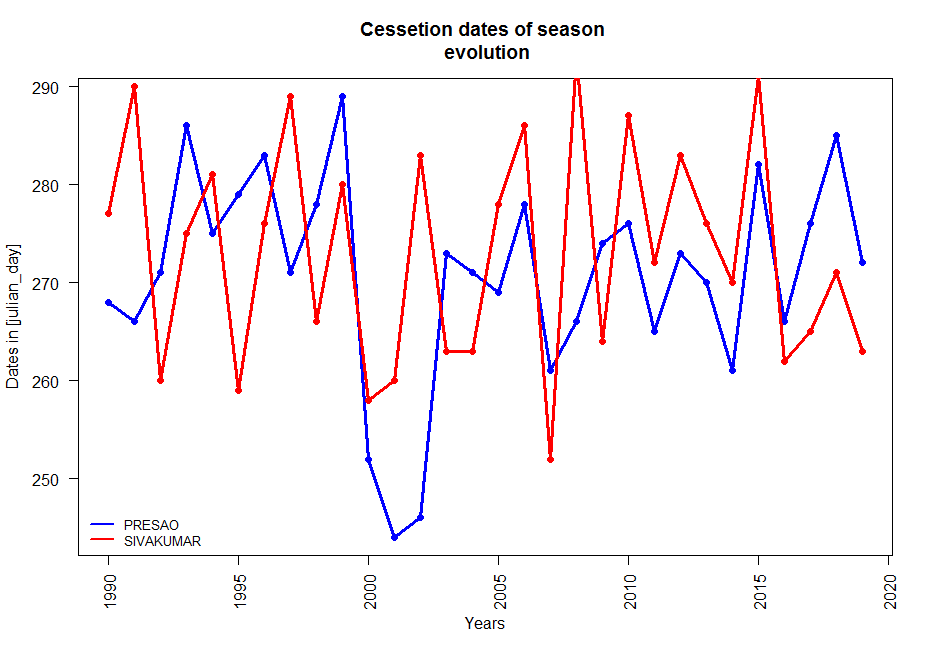}
	
		Figure 19: Cessetion dates variabilities
		\label{Cessetion dates variabilities}
	\end{center}
	Overall, the end of season dates calculated with the SIVAKUMAR criterion are slightly higher than those obtained by the PRESAO criterion. Indeed, 3 years out of 4, the dates for SIVAKUMAR are lower than October 11 and those of PRESAO are lower than October 06 (Figure 20).
	\begin{center}
		\includegraphics[
		height=3.8891in,
		width=4.175in
		]
		{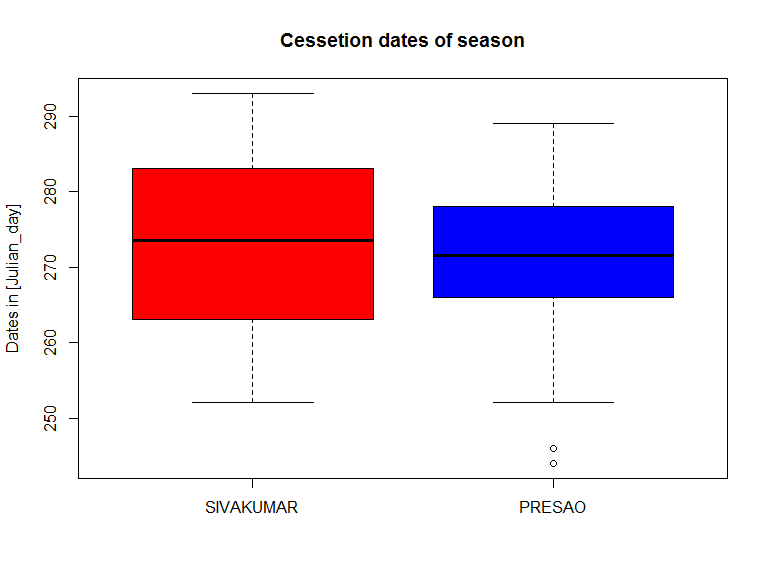}
	
		Figure 20: Cessetion dates distribution
		\label{Cessetion dates distribution}
	\end{center}
	\subsubsection{Analysis of the length of the rainy season}
	
	\bigskip 
	In rainfed cultivation, knowing the length of the rainy season in an area is very crucial in order to make the right decision regarding the variety choice for the success of a given crop. Thus for the two criteria retained in this study, the lengths of the season vary approximately between 60 and 165 days. For an average season length of 125 days, the use of variety of 100 to 120 days is recommended for the study area (Figure 21).
	\begin{center}
		\includegraphics[
		height=3.0683in,
		width=5.1474in
		]
		{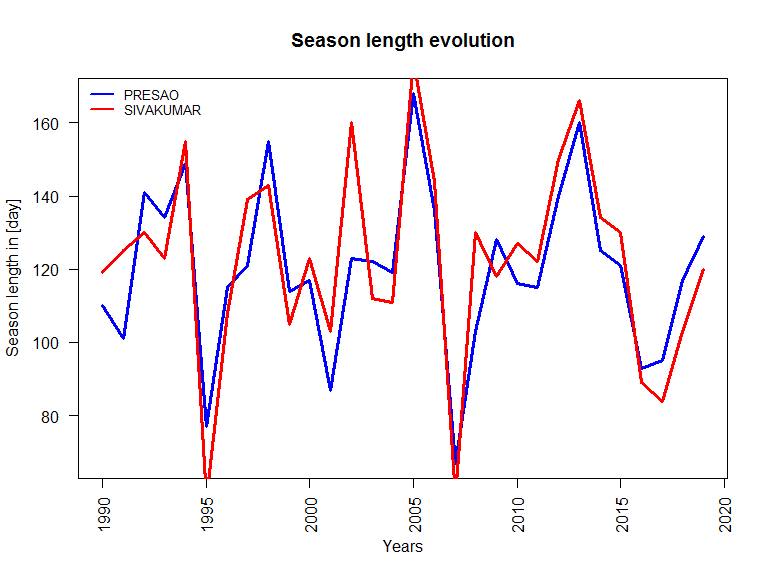}
	
		Figure 21: Season length variabilities
		\label{Season length variabilities}
	\end{center}

	\bigskip 
	For both criteria, the tendency for long or short season lengths is very variable. Nevertheless, the tendency of long seasons seems more marked for the PRESAO criterion (Figure 22).
	\begin{center}
		\includegraphics[
		height=3.8891in,
		width=5.175in
		]
		{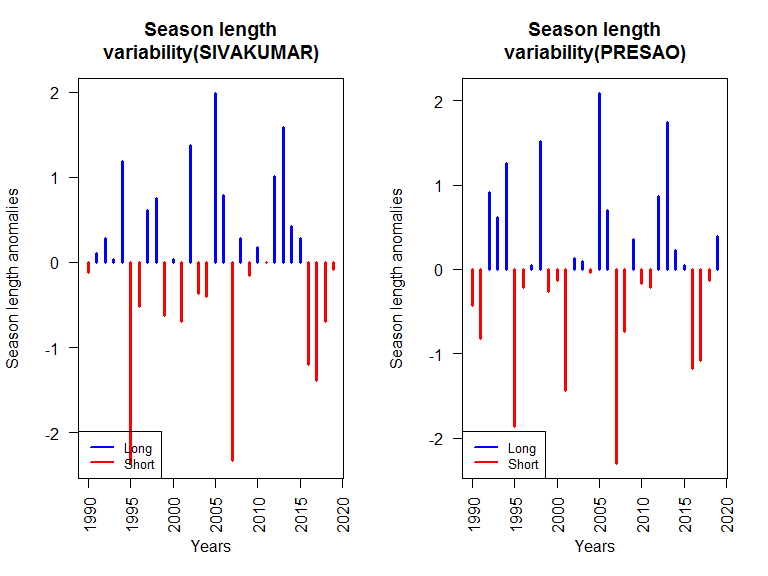}

		Figure 22: Season length anomalies
		\label{Season length anomalies}
	\end{center}
	\subsubsection{Analysis of the number of rainy days in the season}
	
	\bigskip 
	The number of rainy days in the season varies between 20 and 40 days except in extreme cases against 40 and 60 days in the year (Figure 23). Thus, the number of rains on the fringes of the rainy season is less important in the area.
	\begin{center}
		\includegraphics[
		height=3.8891in,
		width=5.175in
		]
		{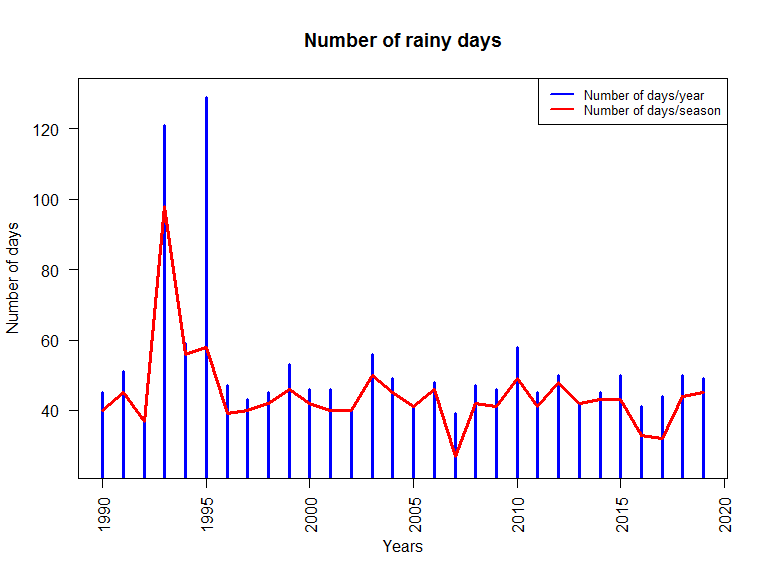}

		Figure 23: Rainy dates evolution
		\label{Rainy dates evolution}
	\end{center}
	\subsection{Correlation analysis}
	
	\subsubsection{Agro-climatic parameters correlation}
	
	\bigskip 
	In general, variables of the same type show a strong correlation with each other (Figure 24). Indeed, the groups of variables such as (Tmax, Tmin, Tmean) and (Umax, Umin) naturally exhibit a strong positive correlation. On the other hand, it is important to underline certain more or less strong connections but in the opposite direction. This is the case, for example, with maximum temperature and relative humidity. Thus, hotter it is, less is water vapor in the atmospheric air. The inverse correlation between minimum
	temperature and insolation can be explained by the fact that the minimum temperature is recorded at 06h UT and therefore before the appearance of sunlight. The negative correlation between maximum temperature and rain is explained by the fact that the occurrence of precipitation reduces temperatures in general. The paradox with the minimum temperatures and the rain at this station could be explained by the fact that most of the rains occur during the day and therefore have more impact on the maximum temperatures than the minimum. Wind and relative humidity correlate in the opposite direction since the wind is a factor of evaporation and therefore participates in the drying of air masses.
	\begin{center}
		\includegraphics[
		height=3.4891in,
		width=5.175in
		]
		{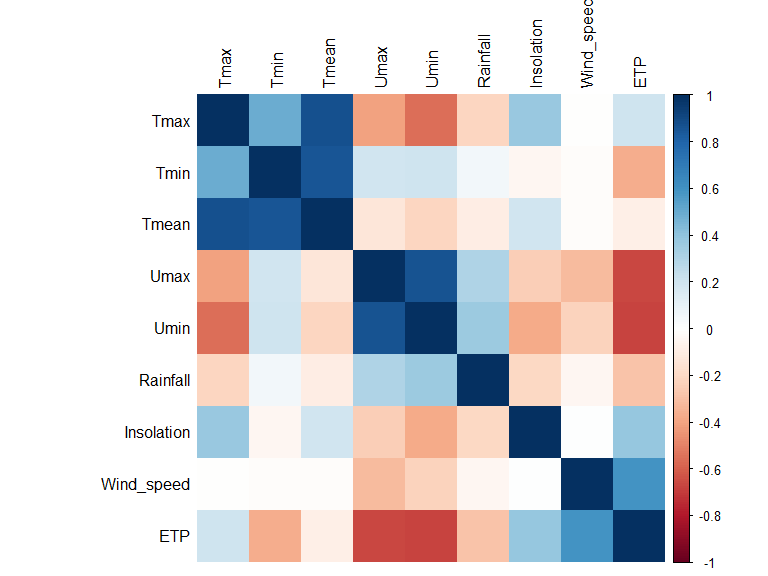}
	
		Figure 24: Ago-climatic parameters correlation
		\label{Ago-climatic parameters correlation}
	\end{center}

	\subsubsection{Characteristic parameters of the season correlation}
	\bigskip
	In general, there is a significantly negative correlation between the onset date and the other parameters characteristic of the rainy season (Figure 25). A result which corroborates the results of previous studies as in \cite{sivakumar1988predicting}. Moreover, the PRESAO method for the end of season date seems to contradict this assertion on our data by showing a significantly positive correlation with the onset date. This last result justifies our choice for the Sivakumar method in the rest of our study.
	\begin{center}
		\includegraphics[
		height=4.0811in,
		width=4.3811in
		]
		{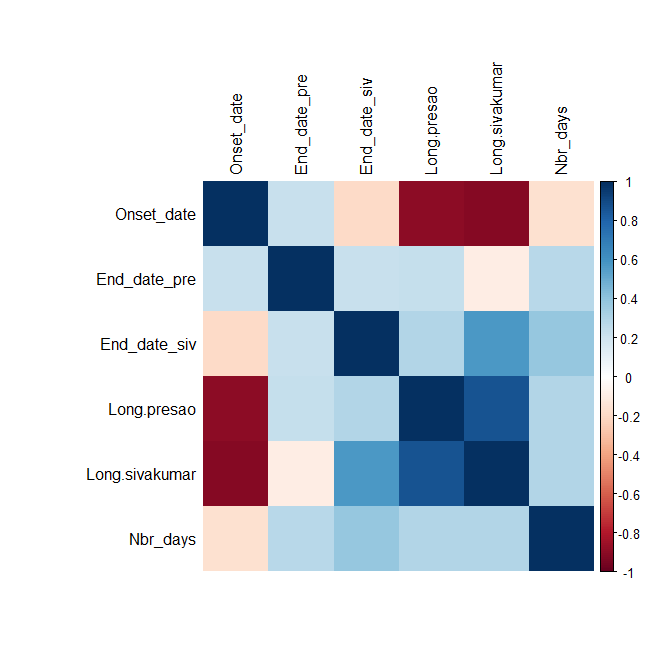}
		
		Figure 25: Caracteristic parameters of season correlation
		\label{Caracteristic parameters of season correlation}
	\end{center}

	\subsection{Identification of the probability distributions of agro-climatic risks}
	\bigskip
	The normal law seems to be the law of probability which best adjusts the various key parameters of the rainy season (onset and cessetion date of the season, length of the season). Indeed, the Shapiro-Wilks tests and the quantile-quantile plots lead to the same conclusions in general.
	
	\subsubsection{Onset dates probability distribution}
	
	\begin{center}
		\includegraphics[
		height=3.8891in,
		width=5.175in
		]
		{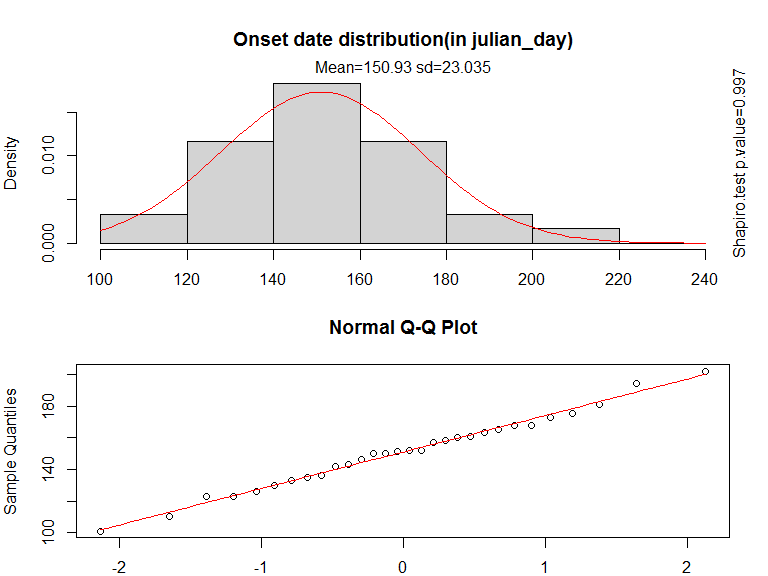}
	
		Figure 26: Onset dates adequation
		\label{Onset dates adequation}
	\end{center}

	\subsubsection{Length of the season probability distribution}
	
	\paragraph{Sivakumar criterion}
	
	\bigskip
	
	\bigskip

	\begin{center}
		\includegraphics[
		height=3.8891in,
		width=5.175in
		]
		{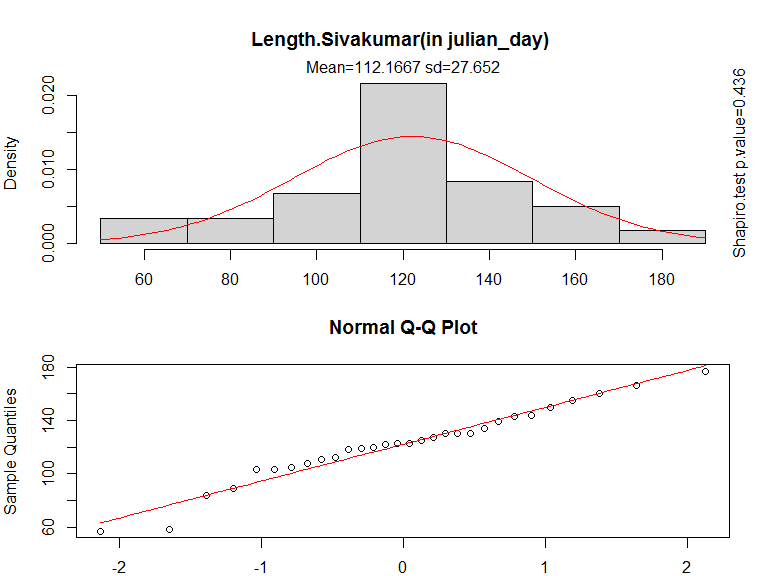}
	
		Figure 27: Length of season adequation (Sivakumar)
		\label{Length of season adequation (Sivakumar)}
	\end{center}

	\paragraph{PRESAO criterion}
	
	\bigskip
	
	\begin{center}
		\includegraphics[
		height=3.8891in,
		width=5.175in
		]
		{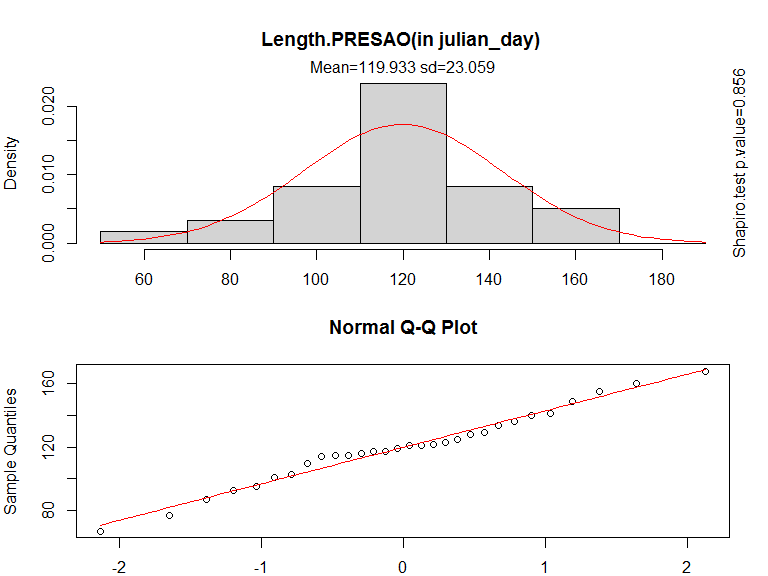}
	
		Figure 28: Length of season adequation (PRESAO)
		\label{Length of season adequation (PRESA)}
	\end{center}
	
	\subsubsection{Cessetion dates probability distribution}
	
	\bigskip
	
	\paragraph{Sivakumar criterion}
	
	\bigskip
	
	\bigskip
	\begin{center}
		\includegraphics[
		height=3.8891in,
		width=5.175in
		]
		{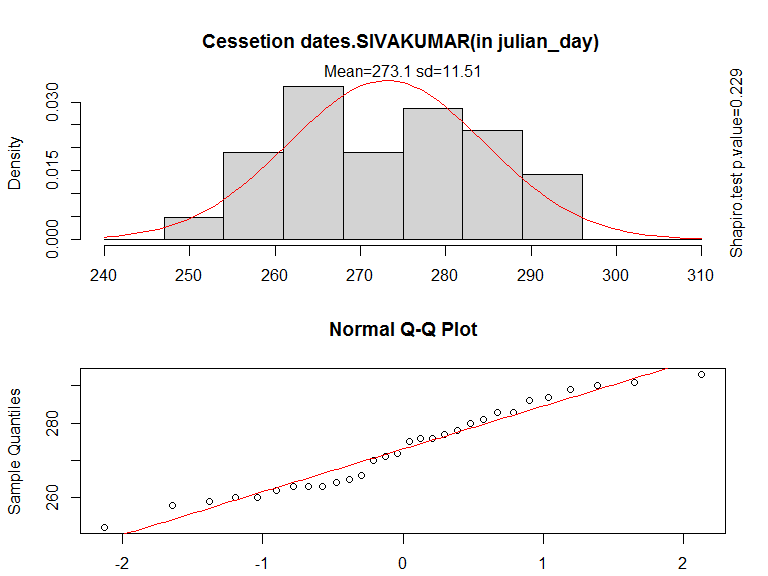}
	
		Figure 29: Cessetion dates adequation (Sivakumar)
		\label{Cessetion dates adequation (Sivakumar)}
	\end{center}

	\paragraph{PRESAO criterion}
	
	\bigskip
	\begin{center}
		\includegraphics[
		height=3.8891in,
		width=5.175in
		]
		{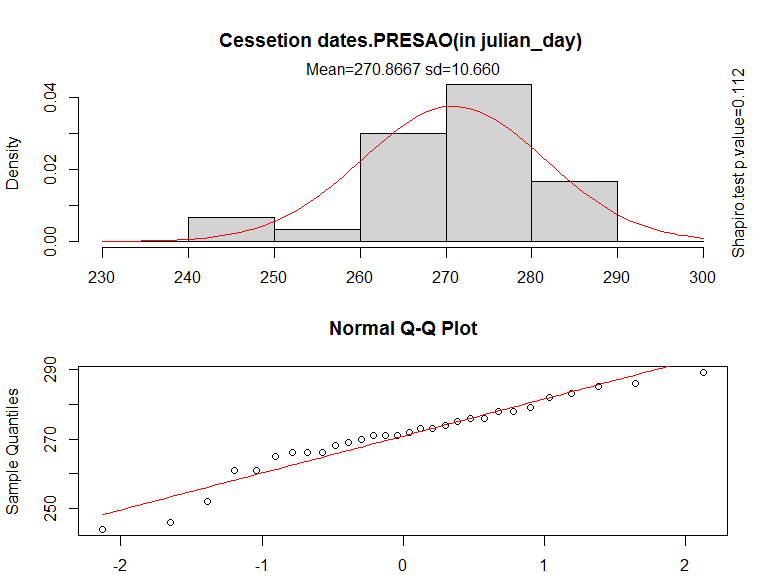}
		
		Figure 30: Cessetion dates adequation (PRESAO)
		\label{Cessetion dates adequation (PRESAO)}
	\end{center}
	\subsubsection{Dry sequences distribution}
		\bigskip
	Over a period of 30 years, the most recurrent dry sequences in the area have a length less than or equal to 7 days (Figure 31). They constitute 93\% of the dry sequences observed. However, dry sequences of greater duration are also observed. This last result is probably due to the false start observed over
	the years 2002, 2005, 2013 and 2019.
	\begin{center}
		\includegraphics[
		height=3.2448in,
		width=3.8891in
		]
		{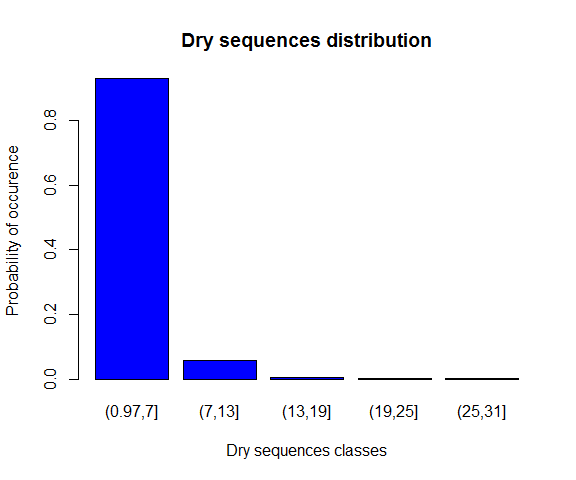}
		
		Figure 31: Dry sequences distribution
		\label{Dry sequences distribution}
	\end{center}

	\section{Conclusion}
		\bigskip
	This study allowed us to contribute to the analysis of the evolution of agro-climatic risks in a context of climate variability in the region of S\'{e}gou in Mali. So, we have used statistical tests on variables
	characterizing the rainy season and we identified the trends and variabilities contained in the data series. We have proposed a statistical modeling of these different agro-climatic risks. Tthe seasonal variability of agro-climatic parameters (temperature, rainfall , humidity, wind speed and insolation) have
	been analized as well as their inter annual variability. Furthermore the probability distributions of agroclimatic risks have been identified and the characterization of the rainy season was clarified.
	
	\bigskip
	
	\bigskip
	\bibliographystyle{abbrvnat}
	\bibliography{Mesreferences}

\begin{thebibliography}{17}
\providecommand{\natexlab}[1]{#1}
\providecommand{\url}[1]{\texttt{#1}}
\expandafter\ifx\csname urlstyle\endcsname\relax
  \providecommand{\doi}[1]{doi: #1}\else
  \providecommand{\doi}{doi: \begingroup \urlstyle{rm}\Url}\fi

\bibitem[Assemian et~al.(2013)Assemian, Kouame, Djagoua, Affian, Jourda, Adja,
  Lasm, and Biemi]{assemian2013etude}
E.~Assemian, F.~Kouame, {\'E}.~Djagoua, K.~Affian, J.~P.~R. Jourda, M.~Adja,
  T.~Lasm, and J.~Biemi.
\newblock {\'E}tude de l'impact des variabilit{\'e}s climatiques sur les
  ressources hydriques d'un milieu tropical humide: cas du d{\'e}partement de
  bongouanou (est de la c{\^o}te d'ivoire).
\newblock \emph{Revue des sciences de l'eau/Journal of Water Science},
  26\penalty0 (3):\penalty0 247--261, 2013.
\newblock \doi{10.7202/1018789ar}.

\bibitem[Bazzaz and Sombroek(1997)]{bazzaz1997changements}
F.~Bazzaz and W.~Sombroek.
\newblock Changements du climat et production agricole.
\newblock \emph{FAO et Polytechnica, pp1-10}, 1997.

\bibitem[Bell and Lamb(2006)]{Lamb2006}
M.~A. Bell and P.~J. Lamb.
\newblock Integration of weather system variability to multidecadal regional
  climate change: The west african sudan-sahel zone, 1951-98.
\newblock \emph{Journal of Climate}, 19\penalty0 (20):\penalty0 5343--5365,
  2006.
\newblock ISSN 08948755, 15200442.
\newblock \doi{https://doi.org/10.1175/JCLI4020.1}.

\bibitem[Beno{\^\i}t(2008)]{benoit2008changements}
{\'E}.~Beno{\^\i}t.
\newblock Les changements climatiques: vuln{\'e}rabilit{\'e}, impacts et
  adaptation dans le monde de la m{\'e}decine traditionnelle au burkina faso.
\newblock \emph{VertigO-la revue {\'e}lectronique en sciences de
  l'environnement}, 8\penalty0 (1), 2008.
\newblock \doi{https://doi.org/10.4000/vertigo.1467}.

\bibitem[Berg et~al.(2013)Berg, de~Noblet-Ducoudr{\'e}, Sultan, Lengaigne, and
  Guimberteau]{berg2013projections}
A.~Berg, N.~de~Noblet-Ducoudr{\'e}, B.~Sultan, M.~Lengaigne, and
  M.~Guimberteau.
\newblock Projections of climate change impacts on potential c4 crop
  productivity over tropical regions.
\newblock \emph{Agricultural and Forest Meteorology}, 170:\penalty0 89--102,
  2013.
\newblock \doi{https://doi.org/10.1016/j.agrformet.2011.12.003}.

\bibitem[DIOUF et~al.(2001)DIOUF, NONGUIERMA, AMANI, and Royer]{diouf2001lutte}
M.~DIOUF, A.~NONGUIERMA, A.~AMANI, and A.~Royer.
\newblock Lutte contre la s{\'e}cheresse au sahel: r{\'e}sultats, acquis et
  perspectives au centre r{\'e}gional agrhymet.
\newblock \emph{Science et changements plan{\'e}taires/S{\'e}cheresse},
  11\penalty0 (4):\penalty0 257--66, 2001.
\newblock ISSN 1147-7806.

\bibitem[Le~Barb{\'e} et~al.(2002)Le~Barb{\'e}, Lebel, and
  Tapsoba]{leBarbe2002rainfall}
L.~Le~Barb{\'e}, T.~Lebel, and D.~Tapsoba.
\newblock Rainfall variability in west africa during the years 1950--90.
\newblock \emph{Journal of climate}, 15\penalty0 (2):\penalty0 187--202, 2002.
\newblock
  \doi{https://doi.org/10.1175/1520-0442(2002)015<0187:RVIWAD>2.0.CO;2}.

\bibitem[Lubes et~al.(1994)Lubes, Masson, Servat, Paturel, Kouame, and
  Boyer]{lubes1994caracterisation}
H.~Lubes, J.~Masson, E.~Servat, J.~Paturel, B.~Kouame, and J.~Boyer.
\newblock Caract{\'e}risation de fluctuations dans une s{\'e}rie chronologique
  par application de tests statistiques.
\newblock \emph{Etude bibliographique, Programme ICCARE, Rapport}, 3:\penalty0
  1--21, 1994.
\newblock URL \url{https://www.documentation.ird.fr/hor/fdi:010020654}.

\bibitem[Mann(1945)]{mann1945nonparametric}
H.~B. Mann.
\newblock Nonparametric tests against trend.
\newblock \emph{Econometrica: Journal of the econometric society}, pages
  245--259, 1945.
\newblock \doi{https://doi.org/10.2307/1907187}.
\newblock URL \url{https://www.jstor.org/stable/1907187}.

\bibitem[Moorcroft et~al.(2006)Moorcroft, Pacala, and
  Lewis]{moorcroft2006potential}
P.~Moorcroft, S.~W. Pacala, and M.~Lewis.
\newblock Potential role of natural enemies during tree range expansions
  following climate change.
\newblock \emph{Journal of Theoretical Biology}, 241\penalty0 (3):\penalty0
  601--616, 2006.
\newblock \doi{https://doi.org/10.1016/j.jtbi.2005.12.019}.

\bibitem[Nicholson(2001)]{Nicholson2001}
S.~Nicholson.
\newblock Climatic and environmental change in africa during the last two
  centuries.
\newblock \emph{Climate Research - CLIMATE RES}, 17:\penalty0 123--144, 08
  2001.
\newblock \doi{10.3354/cr017123}.

\bibitem[Roudier et~al.(2012)Roudier, Sultan, Quirion, Baron, Alhassane,
  Traor{\'e}, and Muller]{Roudier2012}
P.~Roudier, B.~Sultan, P.~Quirion, C.~Baron, A.~Alhassane, S.~B. Traor{\'e},
  and B.~Muller.
\newblock An ex-ante evaluation of the use of seasonal climate forecasts for
  millet growers in sw niger.
\newblock \emph{International Journal of Climatology}, 32\penalty0
  (5):\penalty0 759--771, 2012.
\newblock \doi{https://doi.org/10.1002/joc.2308}.
\newblock URL
  \url{https://rmets.onlinelibrary.wiley.com/doi/abs/10.1002/joc.2308}.

\bibitem[Sivakumar(1988)]{sivakumar1988predicting}
M.~Sivakumar.
\newblock Predicting rainy season potential from the onset of rains in southern
  sahelian and sudanian climatic zones of west africa.
\newblock \emph{Agricultural and forest meteorology}, 42\penalty0 (4):\penalty0
  295--305, 1988.
\newblock \doi{https://doi.org/10.1016/0168-1923(88)90039-1}.

\bibitem[Sivakumar(1992)]{Sivakumar1992}
M.~V.~K. Sivakumar.
\newblock Empirical analysis of dry spells for agricultural applications in
  west africa.
\newblock \emph{Journal of Climate}, 5\penalty0 (5):\penalty0 532--539, 1992.
\newblock ISSN 08948755, 15200442.
\newblock
  \doi{https://doi.org/10.1175/1520-0442(1992)005<0532:EAODSF>2.0.CO;2}.
\newblock URL \url{http://www.jstor.org/stable/26197057}.

\bibitem[Solomon et~al.(2007)Solomon, Manning, Marquis, Qin,
  et~al.]{solomon2007climate}
S.~Solomon, M.~Manning, M.~Marquis, D.~Qin, et~al.
\newblock \emph{Climate change 2007-the physical science basis: Working group I
  contribution to the fourth assessment report of the IPCC}, volume~4.
\newblock Cambridge university press, 2007.

\bibitem[Sultan(2012)]{sultan2012global}
B.~Sultan.
\newblock Global warming threatens agricultural productivity in africa and
  south asia.
\newblock \emph{Environmental Research Letters}, 7\penalty0 (4):\penalty0
  041001, 2012.
\newblock \doi{10.1088/1748-9326/7/4/041001}.
\newblock URL \url{https://doi.org/10.1088/1748-9326/7/4/041001}.

\bibitem[Sultan et~al.(2005)Sultan, Baron, Dingkuhn, Sarr, and
  Janicot]{sultan2005agricultural}
B.~Sultan, C.~Baron, M.~Dingkuhn, B.~Sarr, and S.~Janicot.
\newblock Agricultural impacts of large-scale variability of the west african
  monsoon.
\newblock \emph{Agricultural and forest meteorology}, 128\penalty0
  (1-2):\penalty0 93--110, 2005.
\newblock \doi{https://doi.org/10.1016/j.agrformet.2004.08.005}.

\end{thebibliography}
\end{document}